# Stable Oleoplaned Slippery Surfaces on Biomimetically Patterned Templates

Saumyadwip Bandyopadhyay[a], Sriram S M[b], Anuja Das[c], Rabibrata Mukherjee[c], Suman Chakraborty[a,d]

[a]Advanced Technology Development Centre, Indian Institute of Technology Kharagpur, Kharagpur- 721 302, West Bengal, India.

[b]Department of Mechanical Engineering, National Institute of Technology Karnataka, Surathkal- 575025, Karnataka, India.

[c]Instability and Soft Patterning Laboratory, Department of Chemical Engineering, Indian Institute of Technology Kharagpur, Kharagpur- 721302, West Bengal, India.

[d]Department of Mechanical Engineering, Indian Institute of Technology Kharagpur, Kharagpur- 721302, West Bengal, India.

Abstract

With the advent of the technology of the oleoplaned slippery surfaces as the better solution to self-cleaning, anti fouling and self-healing smart surfaces, the stability of the oil layer on the surfaces has caught a great deal of attention from the research community. Rose petals irrespective of its superhydrophobic nature exhibits a very high adhesion owing to the hierarchical structures and can thus serve as an excellent surface to obtain a stable oil film. Also, with gradual covering of the rose petal structures by the oil the change in the adhesion force is observed to decrease and an increase in the film thickness beyond a certain height causes cloaking of the droplet and thus presents us with an optimum thickness which can give us a stable oil film and also exhibit high degree of slipperiness. The findings can be applied for further applications in droplet based microfluidics, as a low energy actuation surface, or as a self-healing and self-cleaning surface.

Introduction

Starting from public health, to medical equipments and from automobiles to heavy machineries self cleaning surfaces is a subject of absolute importance, as the self-cleaning surfaces hold solution to a multitude of problems ranging from anti-fouling, anti-icing, anti fogging, and anti-corrosive surfaces. Many researchers have suggested the lotus leaf effect inspired structural superhydrophobic surfaces to be a plausible solution to the self-cleaning surfaces[1–7]. However their functionality falls short of expectations under extreme temperatures and high humidity conditions. To this end the ability to maintain a stable air film by these superhydrophobic surfaces is compromised, and the vapors come in contact with the solid surface directly and hence the functionality of these structural superhydrophobic surfaces is lost. Also, due to mechanical wear and tear the structures on the surface lose their ability to maintain their original aspect ratio which is a quintessential to repel a liquid of given surface tension[8,9]. Also, fabrication of such kind of structures on any type of material is a very challenging task. To evade all these shortcomings of a structural superhydrophobicity as a robust self-cleaning surface, researchers have developed a new kind of self-cleaning surface called Liquid infused slippery surfaces abbreviated as LISS[10]. Like the previous case this also derives its origin from a plant, called *Nepenthes Alata*, an insect-eating carnivorous plant[10,11]. The plant relies on its slippery coating to make the insects slip further into the flower and traps it inside[12]. Since then the LISS has gained the attention in many research groups and have been shown to exhibit many qualities like anti-biofouling[11,13–16], anti-corrosive[15,17,18], anti-frost[19], anti-icing[20,19], optically tunable[21], and self-healing[22–24] as well. However, the major advantage of this surface lies in its fabrication, which is it can be fabricated with different kinds of material with not much difficulty. So far, it has been shown to be fabricated on polymers[11,16,21,5], silicon[5,25,26], metals[15,17,18], and glass.

The liquid infused slippery surfaces relies on a stable liquid film covering the solid surfaces to repel water or any other fouling liquid instead of an air film as in the case of a structural superhydrophobic surface. The stability of the liquid infused surface lies on three major conditions. First is the ability of the oil or the lubricating liquid to wet the surface. Second is the immiscibility of the lubricating film with the fouling liquid. Third one is that the solid surface should have a higher affinity towards the lubricating liquid rather than the fouling liquid. So far these liquid infused slippery surfaces have been made by majorly two different methods. One by infusing the lubricating liquid in a porous matrix of a solid media, and second is by imparting roughness to a surface, treating the surface to make it repelling to the fouling liquid and then coating it with oil. Thus it is the latter technique of the fabrication that makes the fabrication of these surfaces much more luring because of its applicability to different kinds of materials. Also, studies have been carried out on the structures from the perspective of a stable liquid film[27]. It was found that a roughness of the order of nanometers can have better stability of the liquid film than micrometer scale roughness. However a hierarchical order roughness can also act as a good anchor for the lubricating film[28].

In nature we find these hierarchical structures on may plant surfaces like the lotus leaf and the rose petals. Due to this advanced structuring on their surfaces water droplets exhibit phenomenally high contact angles on both these surfaces[29]. However the difference lies in the fact that on the lotus leaf the water droplet rolls off with the slightest slant even less than 2°, whereas, on the rose petal the droplet is anchored[29]. This observation has gained much attention because of its uniqueness and applications in different fields[30–33]. Unlike the nanoscale hair like structures interspersed on the micropapillae of the surface on lotus leaf the rose petals have ridge like structures interspersed over the micropapillae. This difference in the structure may be the reason for the easy penetration of the droplets into the nanostructure. Once the droplet impregnates into the nano-ridge like structures it gets pinned. Thus we believe that for developing a surface with a stable liquid film the structures on the rose petal surface can prove beneficial.

In this work we provide a very facile and novel method of fabricating a biomimetic oleoplaned slippery surface. We used the rose petal replica as the base for anchoring the lubricating film. The rose petal replica has been fabricated using a facile two step soft lithographic technique. We show that the effect of thickness and viscosity of the lubricating oil plays a vital role in deciding the adhesion force on these surfaces. This in turn changes the wetting characteristics. This also enhances the degree of slipperiness of the surface. Also, a comparison of the stability of the liquid film has been carried out between flat PDMS and the rose petal replica based LIS. It was found that the rose petal based oleoplaned surfaces are better at anchoring the lubricating film and also, exhibits lower adhesion force making it a better contender for fabricating a stable oleoplaned surface.

**Experimental Methodology**

Rose petal replicas are prepared by two step soft lithographic technique[34,35]. Sylgard 184 a commercially available PDMS based two part (base elastomer and cross-linking agent) polymer mixed in 10:1 (base elastomer to cross-linker) ratio. The mixture is poured on the freshly collected rose petals from the horticulture section of IIT Kharagpur and carefully stuck in a petri dish. The arrangement is then degassed in a dessicator to remove any entrapped air between the rose petals and the uncured Sylgard 184 mixture. After degassing the whole setup is left in the dessicator for the next 5 days to cure properly. After 5 days the cured Sylgard is peeled off from the rose petals. The structures that we get after this process is the negative replica of the rose petals. This negative replica is then treated in an UV oxidation chamber for 50 minutes to form a silicate layer on the PDMS. This silicate layer is a pre requisite for making the positive replica using the same Sylgard 184, as it creates a barrier layer between the already cured Sylgard and the uncured Sylgard 184 mixture which will be used in replicating the negative structures and create positive replica of the rose petals. To make the positive replica on rigid samples (as required for our next experiments) we used 15 mm x 15mm cut glass slides as the base material. The glass slides were then oxygen plasma treated at 40 Watt for 10 minutes to make them more adhering towards the to be coated Sylgard. After that we spun coated the plasma treated glass slides with the uncured Sylgard 184 at 800 rpm for 90 seconds to obtain an uniform layer thickness of 500µm. On these spun coated glass slides we put the UV treated Sylgard 184 negative rose petal replicas and put it in a convection oven at 90°C for 12 hours to cure. After that we peeled off the negative replicas to obtain the positive replicas on the glass slides. The complete procedure is depicted in the schematic shown in Fig.1.

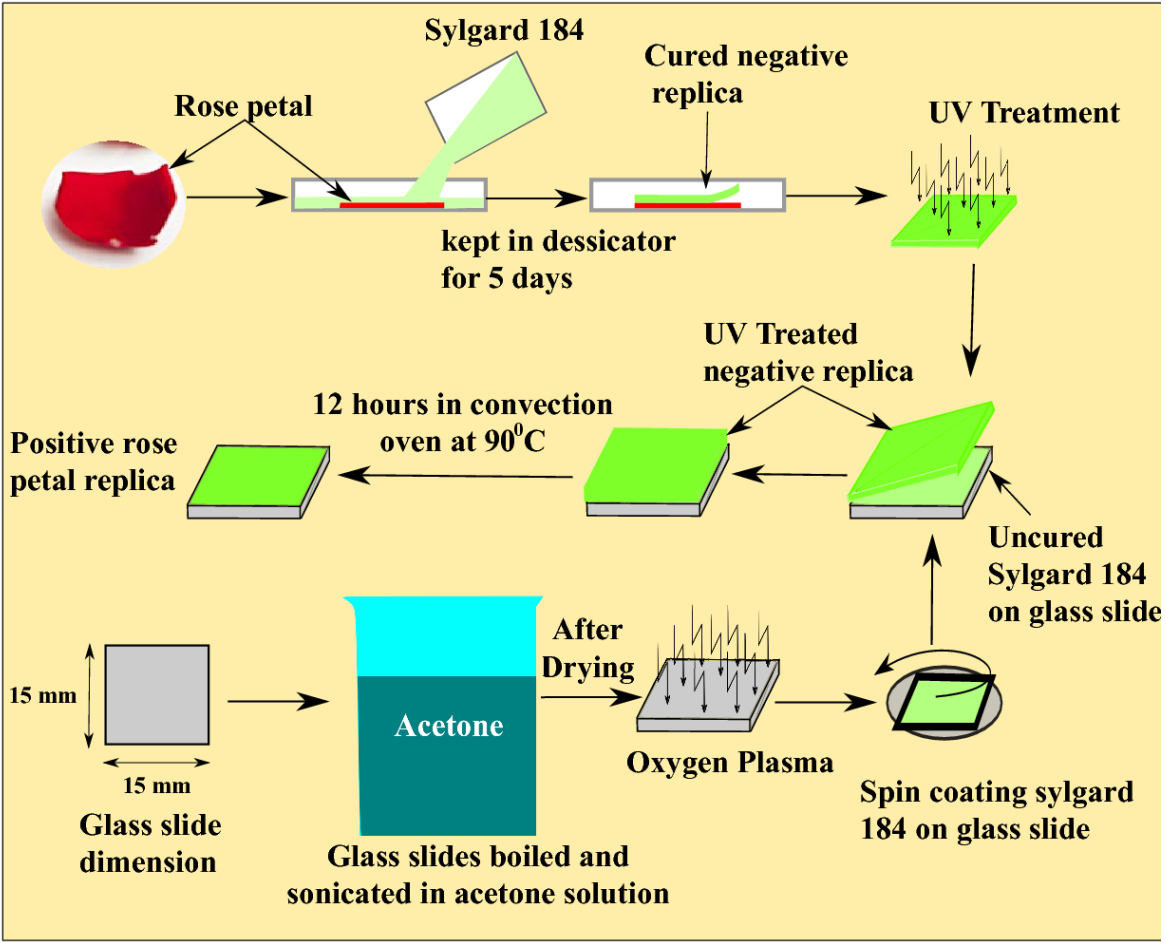

**Fig.1** Schematic of the Sylgard 184 rose petal replica fabrication procedure

The replicas thus obtained from the above mentioned procedure has the same structures as that of the rose petals. The replicas were then again spun coated with Silicone oils of different viscosities ($10^{-4}$, $5\times10^{-4}$, $10^{-3}$, and $10^{-2}$ m2/s) at different rotational speeds to obtain different thicknesses. The speeds were so selected so as to create two thicknesses which are lower than the height of the micro-structures on the rose petals and two thicknesses which are higher than them. Same speeds were used to coat silicone oils on flat PDMS as well. After the samples were coated with the silicone oils, we carried out the adhesion test on them. For adhesion test we used a Laboratory scale chemical weighing balance from Sartorius of 0.01mg accuracy. The sample was placed on the weighing balance and the reading was set to zero by using the tare button. A needle of internal diameter 1.25mm and outer diameter 1.6 mm which is fixed to a screw arrangement equipped syringe that is again held on to a 3-D movable stage, is placed directly on top of the sample surface with a vertical distance of nearly 10mm. The screw arrangement was used to generate a droplet of definite volume, which in our case was 5µl. After carefully generating a pendant droplet of 5 µl the needle was lowered on to the sample surface by using the movement of the 3-D stage. As soon as the droplet touches the surface the reading shows a negative value. The needle is further lowered unless the reading becomes exact zero. After this the needle was raised by 5 µm in each subsequent steps and the reading on the weighing balance is recorded, unless the droplet is either raised with the needle or gets broken in midway, which in either case means the breaking of a physical contact between the needle and the surface[36]. In the cases where the droplet breaks the experiment fails to measure the adhesion force between the droplet and the surface though. However it signifies that the adhesion between that surface and the water droplet is greater than the cohesive force between water molecules.

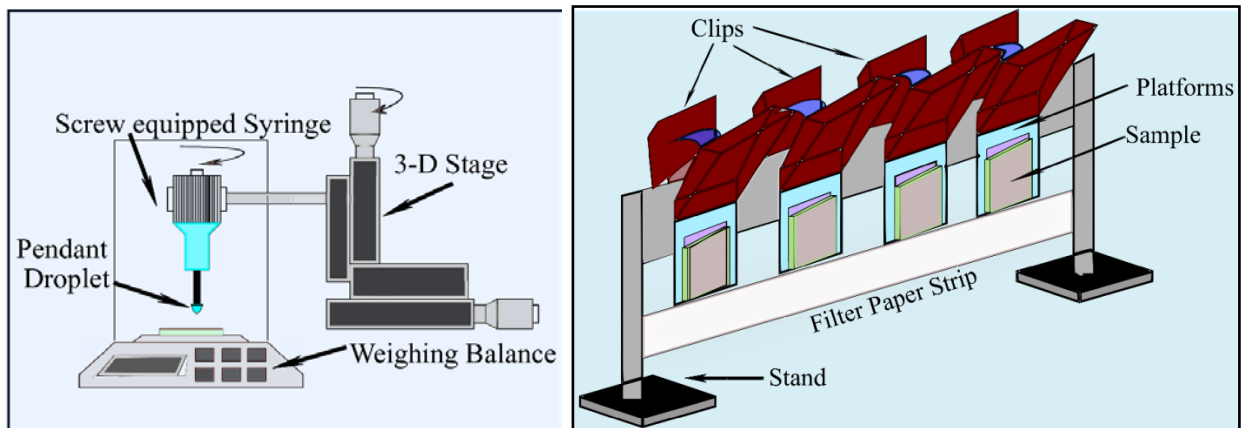

**Fig 2(a).** Schematic of the adhesion test experimental setup. **(b)** Schematic of the oil film draining test setup

The samples were then tested for contact angle and contact angle hysteresis. Also, the lowest tilting angle were measured at which the droplet starts moving on these surfaces. These wetting characterization tests were carried out on a Rame Hart tilting base Goniometer. After complete one day after spin coating the samples with Silicone oils, the test specimens were observed under an optical microscope to examine the stability of the Silicone oil film. Also, a different test was carried out for checking the stability of the Silicone oil film. In this test the samples were mounted on platforms using double sided adhesive tapes. The platforms were then hung vertically to shed off any unanchored Silicone oil. The samples were weighed after every 10 minutes for 60 minutes.

**Results and Discussion**

The rose petal replica as prepared using the above mentioned two step soft lithographic technique is scanned under Atomic force microscopy to study the detailed 3-dimensional nano-scale structures. It is found that the micropapillae which are the bigger structures are of the order of 6.5 µm, and the lengthscale of the smaller ridge like structures is of the order of ~10 nm.

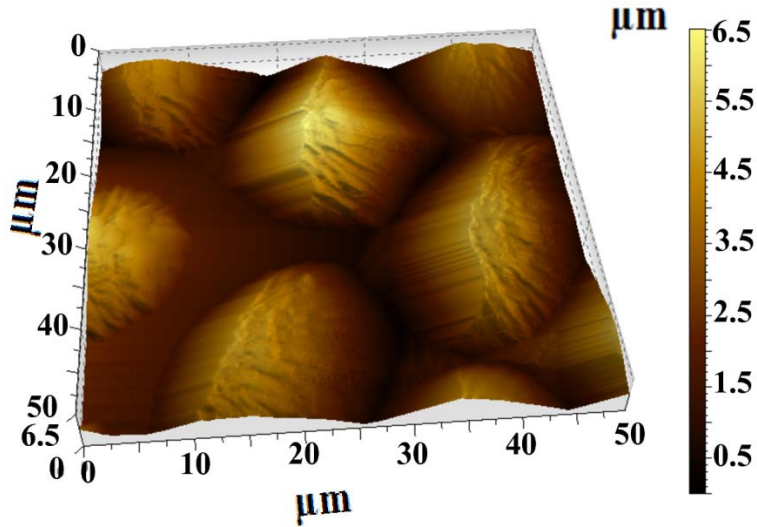

**Fig 3.** AFM scan data of a Sylgard 184 rose petal replica. The height of a papillae is found to be around 6.5μm. The Secondary structures appearing on the micro papillae are the ridge like structures interspersed over the whole surface area.

The thickness, as measured from the change in mass of the sample before and after coating silicone oil and assuming the coating to be uniform all over the surface, is plotted in the graph shown in Fig 3.

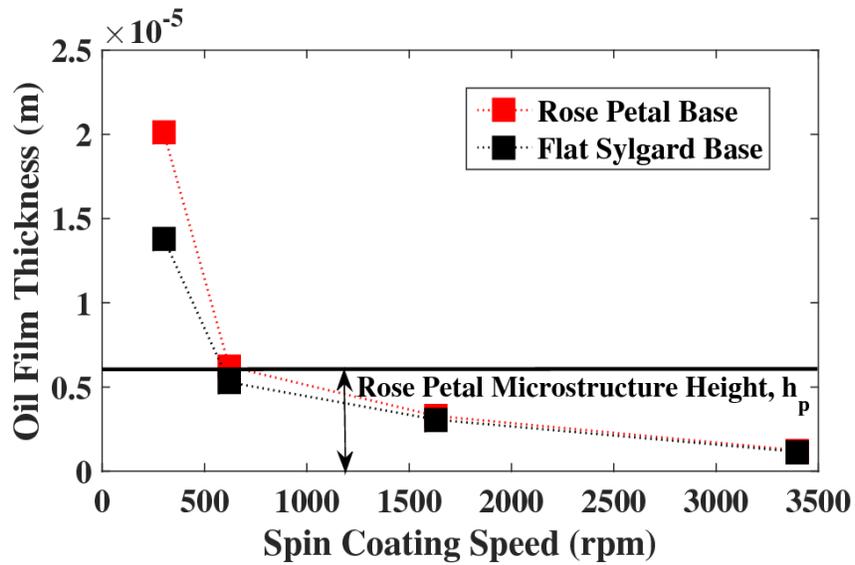

**Fig 4.** Oil thickness with change in spin coating rotational speed for silicone oil of $0.0001 m^2/s$ kinematic viscosity.

As mentioned earlier in the experimental methodology section there are two thicknesses which are less than that of the height of the rose petal surface structures. Also, the height calculated in the above case is by assuming uniform layer, which for flat Sylgard samples may hold correct but never satisfies for rose petal replicas. For rose petal replicas the actual film thickness is greater than the one shown in the above graph, however from the microscopic images shown in Fig 5. we can observe that the broader observation holds good that is the film thickness although greater than the shown value is still less than the height of the microstructures on the rose petal surface. Thus for the two thickness we can see that on the flat Sylgard 184 samples the Silicone oil droplet dewets and forms micro droplets. Which is absent on the rose petals because of the ability to anchor the oil film owing to its nanoscale ridge like structures. It has to be understood that the anchoring of the Silicone oil film is achieved just because of the structures in absence of any chemical modification of the surface. Even for higher thickness oil films we can find that the lubricating liquid is more stable on the rose petal base oleoplaned surfaces than the flat Sylgard 184 surfaces.

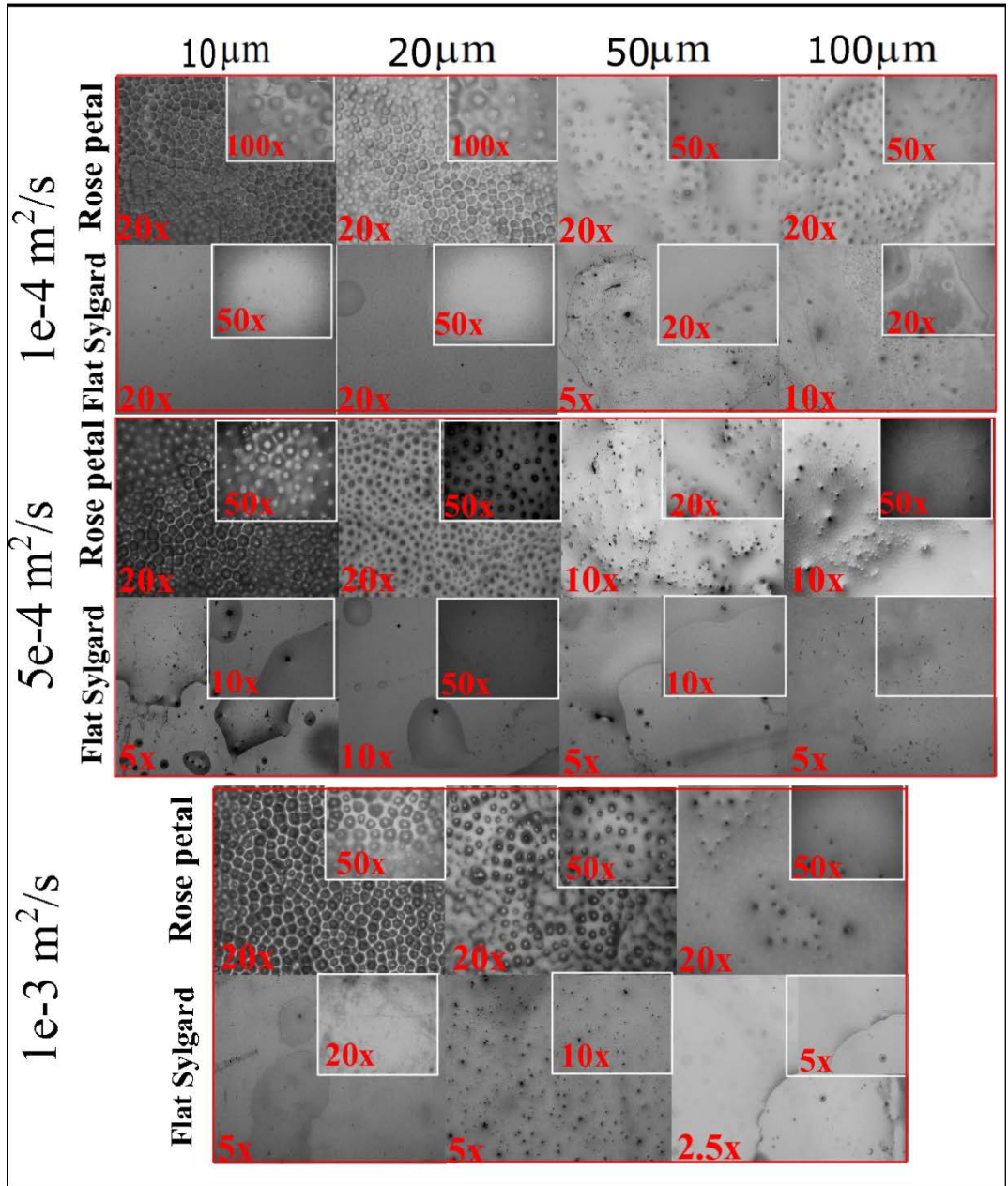

**Fig 5.** Images of the Silicone oil coated surfaces are compared between the rose petal based oleoplaned surfaces and flat Sylgard based oleoplaned surfaces.

The draining of the Silicone oil from the surface is drastically reduced on the rose petal base oleoplaned surfaces as is found from the oil film draining test. The variation of the oil mass fraction with respect to time for two different thickness of oils for up to 1 hour at an interval of 10 minutes is elucidated in the graph in Fig 6.

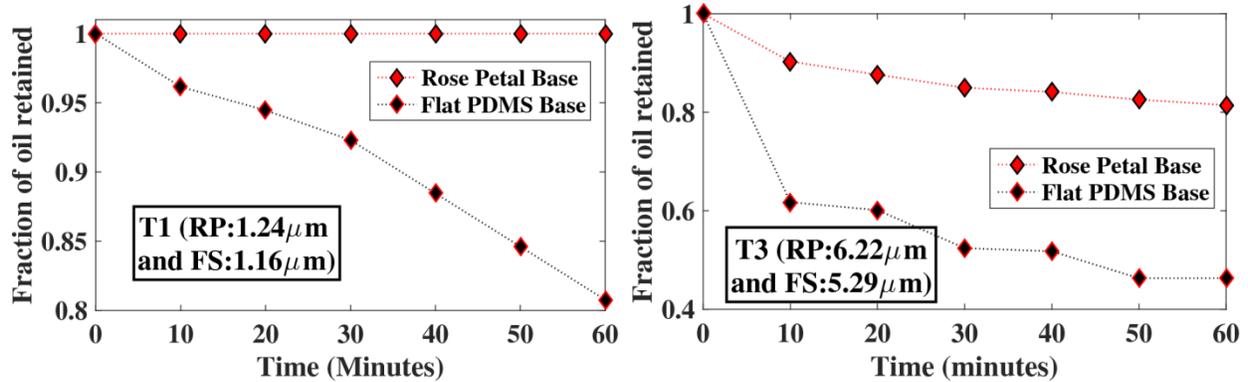

**Fig 6.** The fraction of oil retained on a rose petal based oleoplaned surface and a flat Sylgard 184 oleoplaned surface for two different thickness samples.

On further observation we may understand that the oil film flows out of the rose petal based oleoplaned surface as well, if the oil film thickness is greater than the height of the microstructures on the rose petal surface. Also, from the optical microscope images we can observe that the oil film that is in excess of the microstructure height dewets and forms heap at places.

The effect of this partial covering and complete cloaking of the structures on the rose petal structures also has a significant impact on the adhesive force towards any fouling liquid and thus the wetting state of any other fouling liquid. The adhesive force measurements revealed that as the thickness of the oil film increases the adhesive force between the surface and the fouling liquid (water in this case) increases. This relation is very straight forward in case of the flat Sylgard 184 oleoplaned surfaces, however for rose petal based oleoplaned surfaces the adhesive force at first decreases very gradually and then increases again. This observation is again in coherence with the finding that for the two thicknesses the adhesion force on rose petal based oleoplaned surfaces is higher is because of the partial cloaking of the rose petal structures and thereafter as the thickness of the free film (the film above the rose petal structure height) increases the adhesion starts increasing again, as shown in Fig7.

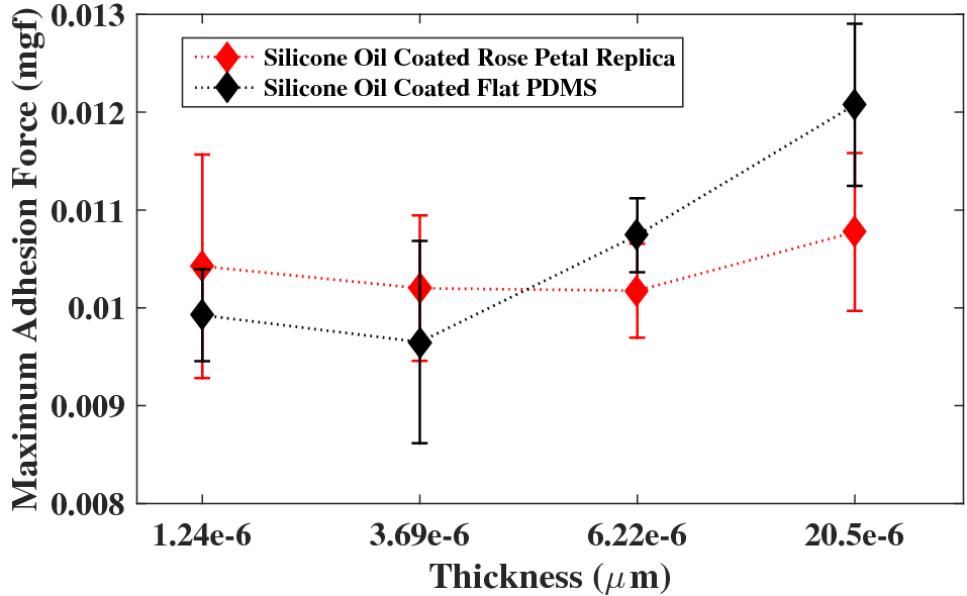

**Fig 7.** Variation of the maximum adhesive force observed for different thickness of the silicone oil film for two different base surfaces (rose petal and flat Sylgard 184). The values plotted are averaged over all the viscosities.

Also, the variation of the maximum force with the viscosity has been studied but the effect is same for both type of surfaces rose petal based and flat Sylgard based. The graph in Fig 8 depicts the variation of the maximum observed adhesive force with the viscosity of the Silicone oil.

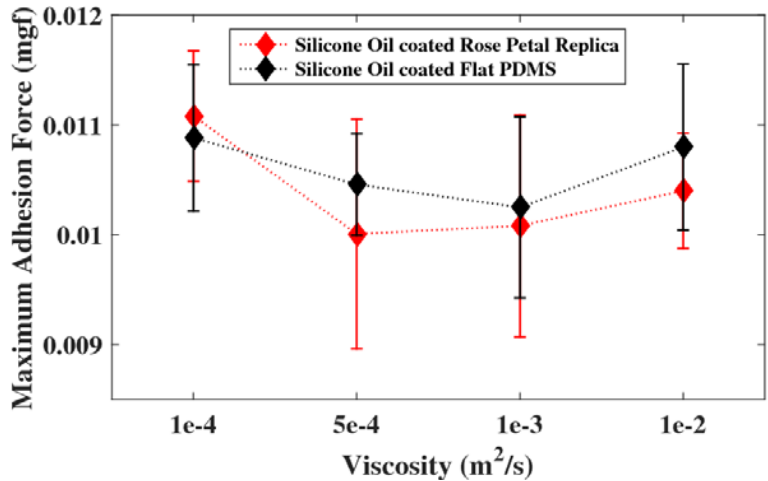

**Fig 8.** Variation of the maximum adhesive force with the viscosity of the lubricating liquid.

The maximum adhesive force decreases and then beyond a certain viscosity (between 5e-4 to 1e-3 $m^2$/s) again starts increasing. This observation is same irrespective of the base surface and thus we assume it to be dependent on the surface property of Sylgard and viscosity of the Silicone oil, and hence is not addressed to in further details in this work as it would diversify the focus at hand.

The effect is also reflected in the contact angle and the dynamic wetting characteristics of the surface. The contact angles for different viscosity silicone oil coated surfaces and of different thicknesses are plotted in Fig 9.

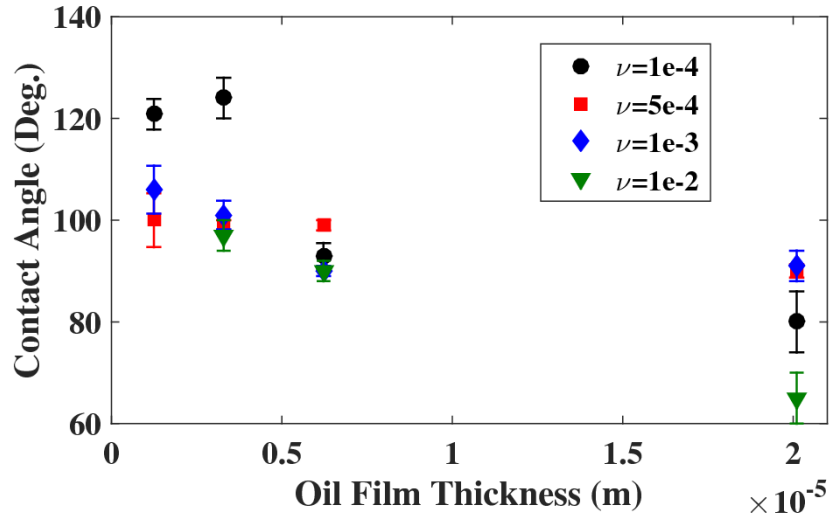

**Fig 9.** Contact angle subtended by a water droplet of 7µl volume on rose petal based oleoplaned surfaces of different conditions. The kinematic viscosity are in m$^2$/s.

The dynamic wetting characteristics have been investigated by studying the two parameters which are namely the contact angle hysteresis (i.e. the difference between the advancing and the receding contact angle at the onset of motion) and the critical tilting angle at which the droplet starts moving (slipping) on the oleoplaned surface. They are plotted in Fig 10(a) and (b) respectively.

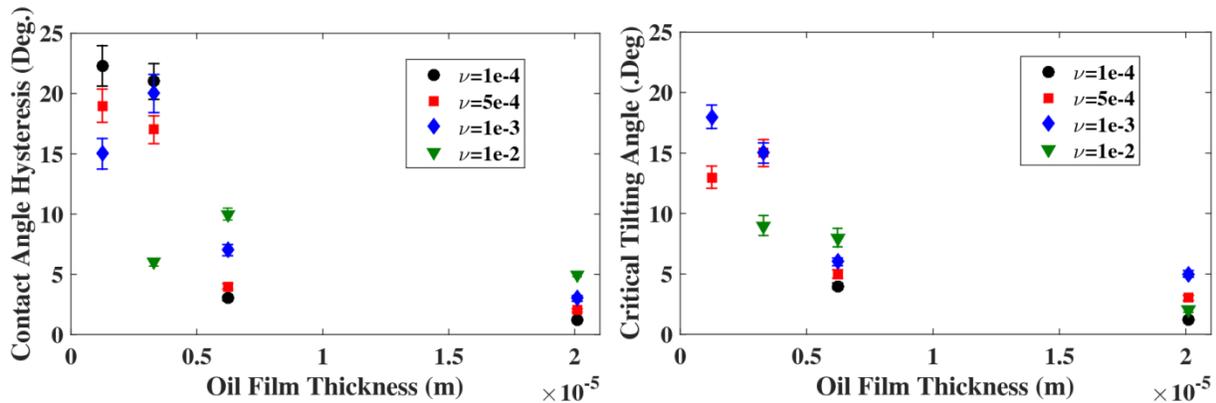

**Fig 10(a).** Variation of contact angle hysteresis for different rose petal based oleoplaned surfaces and in **(b)** variation of the corresponding critical tilting angle with different oil film thickness and different viscosity silicone oil coated rose petal based oleoplaned surfaces.

We can observe that with increase in the thickness of the oil the contact angle hysteresis as well as the critical tilting angle reduces to mere 1°. However the corresponding contact angle is very low and is because cloaking is observed when the film thickness is high. Thus this paradigm is futile from the context of obtaining a stable silicone oil film. It is seen that when cloaking takes place along with the water droplet the lubricating oil also leaves the surface and is hence detrimental to the very cause which is to obtain a stable oil film on the surface.

**Conclusion**

We have been able to demonstrate the ability of a biomimetic surface to anchor a low surface tension and high viscosity liquid by virtue of its surface structures only. The wetting characteristics and the adhesive forces on these kind of surfaces have been studied in great depths. It is found that the adhesive force at first decreases and then increases with increase in the viscosity of the lubricating liquid and monotonously increases with increase in the thickness. However for rose petal based oleoplaned surfaces it deviates because of the incomplete covering of the rose petal structures for low oil film thicknesses which gives rise to direct contact between water and bare rose petal. It can also be concluded from the study that for a certain thickness which is nearly equal (from the greater side) to the structure height of the surface roughness, the surface exhibits maximum degree of slip without exhibiting the phenomenon of cloaking which is detrimental to the functionality of an oleoplaned slippery surface. Thus a stable slippery surface has been developed by a very facile and novel method by applying the adhesive nature of Rose petals in anchoring the lubricating film with greater stability. We believe that this kind of surface can be applied for fabricating self-cleaning, anti-icing, anti-biofouling, and anti-freezing surfaces.